\documentclass[notitlepage,aps,nofootinbib,superscriptaddress,floatfix]{revtex4-1} 
\usepackage{amsmath}
\usepackage{graphicx,epsfig,wrapfig}

\usepackage{amsthm}
\usepackage{makeidx}
\usepackage{amsfonts}
\usepackage{amssymb}
\usepackage{amsmath}
\usepackage{mathtools}
\usepackage[capitalize]{cleveref}
\usepackage{mathrsfs}
\usepackage{slashed}
\usepackage{bm}
\usepackage{color}
\usepackage[normalem]{ulem}
\usepackage{simplewick}

\usepackage{multirow}
\usepackage{braket}

\renewcommand\sout{\bgroup \color[rgb]{0.55,0.00,0.99} \ULdepth=-.5ex \ULset}

\DeclareMathOperator*{\SumInt}{%
\mathchoice%
  {\ooalign{$\displaystyle\sum$\cr\hidewidth$\displaystyle\int$\hidewidth\cr}}
  {\ooalign{\raisebox{.14\height}{\scalebox{.7}{$\textstyle\sum$}}\cr\hidewidth$\textstyle\int$\hidewidth\cr}}
  {\ooalign{\raisebox{.2\height}{\scalebox{.6}{$\scriptstyle\sum$}}\cr$\scriptstyle\int$\cr}}
  {\ooalign{\raisebox{.2\height}{\scalebox{.6}{$\scriptstyle\sum$}}\cr$\scriptstyle\int$\cr}}
}

\begin{document}

\newcommand{\ud}{\text{d}}
\newcommand{\mb}[1]{\boldsymbol{#1}}
\newcommand{\mc}[1]{\mathcal{#1}}
\newcommand{\non}{\nonumber}
\newcommand{\pep}[1]{\mathbf{#1}_{\perp}}
\newcommand{\pepgr}[1]{\bm{#1}_{\perp}}
\newcommand{\gdir}[1]{\gamma^{#1}}
\newcommand{\xk}{(x,\mathbf{k}_{\perp})}
\newcommand{\xkq}{(x,\mathbf{k}_{\perp}^{2})}
\newcommand{\pe}{(\mb{p}_{e})}
\newcommand{\xks}{(x,\pep{k};S)}
\newcommand{\pv}{\mathrm{P.V.}}
\newcommand{\xkl}{(x,\pep{k};\Lambda)}
\newcommand{\dEF}[1]{\nabla^{\text{EF}}_j\left(#1\right)}

\newcommand{\ra}{\rangle}

\newcommand{\di}{\mathrm{d}}
\newcommand{\ta}{\left(}
\newcommand{\qa}{\left[}
\newcommand{\ga}{\left\{}
\newcommand{\tc}{\right)}
\newcommand{\qc}{\right]}
\newcommand{\gc}{\right\}}
\newcommand{\ma}{\left|}
\newcommand{\db}{\right|\left|}
\newcommand{\mnu}[1]{{#1}^{\mu \nu}}
\newcommand{\mnd}[1]{{#1}_{\mu \nu}}
\newcommand{\prodotto}[2]{\ta {#1}\cdot {#2} \tc }
\newcommand{\dq}[1]{\delta^{(4)}\left({#1}\right)}
\newcommand{\dt}[1]{\delta^{(3)} \ta {#1} \tc}
\newcommand{\du}[1]{\delta \ta {#1} \tc}
\newcommand{\vet}[1]{\bm{{#1}}}
\newcommand{\vetp}[1]{\bm{{#1}_{\perp}}}
\newcommand{\pp}{\vet{p}}
\newcommand{\qq}{\vet{q}}
\newcommand{\radice}[1]{\sqrt{{#1}}}
\newcommand{\tr}[1]{\mbox{Tr}\qa {#1} \qc}

\newcommand{\anti}[2]{\overline{{#1}}\ta {#2}\tc }
\newcommand{\antif}[2]{\overline{{#1}}^{(f)}\ta {#2}\tc }
\newcommand{\antim}[2]{\overline{{#1}}_-^{(f)}\ta {#2}\tc }
\newcommand{\antip}[2]{\overline{{#1}}_+^{(f)}\ta {#2}\tc }

\newcommand{\T}{\perp} \newcommand{\Tperp}{T} \newcommand{\bT}{\xi_T}

\newcommand{\note}[1]{\marginpar{\color[rgb]{0.9,0,.9}{\huge$\bullet$}}{\sf\color[rgb]{.9,0,.9}{[#1]}}}
\newcommand{\notebp}[1]{\marginpar{\color[rgb]{0.,0.,1}{\huge$\bullet$}}{\sf\color[rgb]{0,0,1}{[#1]}}}

\title{Revisiting model relations between T-odd  transverse-momentum dependent parton distributions and generalized parton distributions}

\author{Barbara Pasquini}
\email{barbara.pasquini@pv.infn.it}
\affiliation{Dipartimento di Fisica, Universit\`a degli Studi di Pavia, I-27100 Pavia, Italy}
\affiliation{Istituto Nazionale di Fisica Nucleare, Sezione di
  Pavia,  I-27100 Pavia, Italy}

\author{Simone Rodini}
\email{simone.rodini01@ateneopv.it}
\affiliation{Dipartimento di Fisica, Universit\`a degli Studi di Pavia, I-27100 Pavia, Italy}
\affiliation{Istituto Nazionale di Fisica Nucleare, Sezione di
  Pavia,  I-27100 Pavia, Italy}

\author{Alessandro Bacchetta}
\email{alessandro.bacchetta@unipv.it}
\affiliation{Dipartimento di Fisica, Universit\`a degli Studi di Pavia, I-27100 Pavia, Italy}
\affiliation{Istituto Nazionale di Fisica Nucleare, Sezione di
  Pavia,  I-27100 Pavia, Italy}

\date{\today}
\allowdisplaybreaks[2]

\begin{abstract}
We revisit the connection between  generalized parton distributions in impact parameter space and T-odd effects  in single spin asymmetries of the semi-inclusive deep inelastic process.
We show that nontrivial relations can be established
only under very specific conditions, typically realized only
in models that describe hadrons
as two-body bound systems and involving a helicity-conserving
coupling between the gauge boson and the spectator system. Examples of these
models are the
the scalar-diquark spectator  model or the quark-target model for the nucleon,
and relativistic models for the pion at the lowest order in the Fock-space
expansion. 
\end{abstract}

\maketitle
\section*{Introduction}
Generalized parton distributions (GPDs) and transverse momentum dependent parton distributions (TMDs) are fundamental nonperturbative objects  
that help unraveling the quark-gluon dynamics inside  hadrons. 
At leading twist, there are eight independent  GPDs and eight independent TMDs, 
in a one to one correspondence depending on the active parton and target polarizations.
This correspondence arises  from the projection of the fully unintegrated and
off-diagonal correlator, defining  the generalized transverse momentum
dependent parton distributions, into two independent subspaces of the whole space spanned by the parton and target momentum~\cite{Meissner:2008ay,Meissner:2009ww,Lorce:2013pza,Lorce:2011dv,Burkardt:2015qoa}.
Furthermore, one can define impact-parameter dependent densities (IPDs) as the 
 Fourier transforms of the GPDs in impact parameter space at zero longitudinal momentum transfer. The correlator of IPDs
has formally the same structure   as the correlator for TMDs, 
with the impact parameter $\vet{b}_\perp$ taking the role of the transverse momentum $\vet{k}_\perp$~\cite{Diehl:2005jf,Meissner:2007rx}.
Beyond this formal connection, in general it is not possible to establish model-independent relations between GPDs and TMDs.
Only model calculations  show nontrivial relations~\cite{Meissner:2007rx,Avakian:2010br,Lorce:2013pza}.
The most prominent cases are the relations
which describe T-odd effects in single spin asymmetries (SSAs) via factorization of the effects of final state interactions (FSIs), incorporated in a so-called ``chromodynamics lensing function,'' and a spatial distortion of GPDs in impact parameter space~\cite{Burkardt:2002ks,Burkardt:2003uw}.
 These relations have been established for the Sivers effect and the IPD
 for unpolarized partons in a transversely polarized nucleon target,
 using spectator models~\cite{Burkardt:2003je,Bacchetta:2008af}
 and a quark target
  model~\cite{Meissner:2007rx}, and used also in a phenomenological extraction
  of the Sivers function~\cite{Bacchetta:2011gx}.
  Furthermore, they have been discussed for the Boer-Mulders effect and a  certain combination of chiral-odd IPDs describing transversely polarized quark in an unpolarized target, such as the nucleon~\cite{Burkardt:2005hp} or the pion~\cite{Burkardt:2007xm,Gamberg:2009uk,Meissner:2008ay}. 
  However, they have been found to be violated within three-quark model
  calculations for the
  nucleon~\cite{Pasquini:2010af,Pasquini:2008ax,Boffi:2002yy,Pasquini:2005dk,Pasquini:2007xz}. More
  in general, it has been argued that even in the context of spectator models
  these relations  are far from being obvious if one considers Fock-state
  contributions beyond the leading terms~\cite{Meissner:2007rx}. 
 
In this work, we demonstrate that  very specific
conditions have to be imposed on the final-state interaction in order  to express T-odd TMDs in terms of an
impact-parameter distortion and a lensing function.
These conditions are typically
fulfilled only in models where the target is described as two-body bound system and the FSI does not change any of the spectator's quantum numbers and modifies
only its transverse momentum. 
 
The work is organized as follows. In Sec.~\ref{TheoreticalRelations}, we introduce the definition of the  GPDs, both in momentum and  impact parameter space, and of the
 TMDs. We then discuss the possible lensing relations between the average transverse momentum related to T-odd effects and IPDs,
 and the very restrictive
 conditions that should be imposed on the FSI for the validity of these relations. 
 In Sec.~\ref{PionLensingFunction}, we derive the lensing relation for the pion target, described in terms of the lowest $q\bar q$ Fock-state component.
 Besides the two-body nature of the system, a key ingredient for the validity of the relation is the assumption of a perturbative coupling between the spectator parton and the Wilson gluon, which gives the effects of the FSIs in a semi-inclusive deep inelastic scattering (SIDIS) process. The relation can be  generalized  by assuming an effective interaction vertex, which is helicity conserving and may depend only on the transverse-momentum of the exchanged gluon.
  In Sec.~\ref{ProtonLensing}, we consider the case of a proton target and we show that the lensing relations can not be established in a model-independent way 
 even in the most simple case of  a proton target described by the lowest-order Fock component of three quarks.
 Sec.~\ref{ProtonModels} deals with  models that  describe the nucleon as a two-body system, such as  the models with a quark and a diquark spectator, and 
 elucidates under which conditions one can restore the lensing relations
 within these models. Our conclusions are drawn in Sec.~\ref{sect:conclusions}.
 In the Appendix we show the derivation of the conditions that should be satisfied for the lensing relation. 
 
\section{Relations between GPDs and T-odd TMDs}
\label{TheoreticalRelations}

In this section, we summarize  the arguments that lead to infer a possible nontrivial relation between T-odd TMDs  and IPDs.
The quark TMDs are defined through the following correlation function
\begin{align}
\Phi^{[\Gamma]}\ta x, \vetp{k}, S\tc &= \frac{1}{2}\int \frac{dz^-d\vetp{z}}{(2\pi)^3} e^{ik\cdot z}\braket{p,S|\overline\psi\ta -\frac{z}{2}\tc \Gamma \mathcal{W}\ta-\frac{z}{2},\frac{z}{2}\tc \psi\ta \frac{z}{2}\tc|p,S}\Big|_{z^+=0},
\label{TMDCORR} 
\end{align}
where $p=(p^+,p^-,\vet{p}_\perp=\vetp{0})$ and $S$ are, respectively, the hadron-target 
momentum\footnote{We use light-front coordinates, with $v^\pm=1/\sqrt{2}
  (v^0\pm  v^3 )$ and $\vet{v}_\perp=(v^1,v^2)$ for a generic four-vector
  $v$.} 
and spin, $\psi$ is the quark field operator and $\Gamma$ is a generic matrix
in the Dirac space.
The TMDs depend on the light-cone momentum fraction
\begin{equation} 
x=\frac{k^+}{P^+},
\end{equation} 
and on the quark transverse momentum $\vetp{k}$. 

The Wilson line $\mathcal{W}$ connecting the two quark fields  ensures color gauge invariance and is defined as~\cite{Collins:2002kn,Collins:1981uw,Ji:2002aa,Belitsky:2002sm,Boer:2003cm}
\[
\mathcal{W}(a,b) = \mathcal{P}\text{exp}\left\{ -ig_s\int_{\gamma} d\zeta \cdot A(\zeta) \right\},
\]
where $g_s$ is related to the strong coupling constant, $\alpha_s=g^2_s/(4\pi)$, and $\gamma$ is a path from $a$ to $b$ that is determined by the physical process under consideration (in this work, we consider the Wilson line of a SIDIS process)~\cite{Collins:2002kn}. 
 The Wilson line in Eq.~\eqref{TMDCORR}
 breaks the naive time-reversal invariance of the correlator and, as a
 consequence, T-odd TMDs need to be included.
 At leading-twist and for spin $1/2$ targets, one has two T-odd TMDs, i.e. the 
 Sivers function $f_{1T}^{\perp}(x,\vet{k}^2_\perp)$~\cite{Sivers:1989cc,Sivers:1990fh} and the Boer-Mulders TMD $h_1^\perp(x,\vet{k}^2_\perp)$~\cite{Boer:1997nt}.
 The Sivers function describes  the momentum distribution of unpolarized quark in a transversely polarized target, and is obtained from the correlator~\eqref{TMDCORR} with $\Gamma=\gamma^+$.
 On the other hand, 
the Boer-Mulders function gives the momentum distribution of transversely polarized quarks in an unpolarized target, and is 
 is defined from the correlator~\eqref{TMDCORR} with  
 $\Gamma=i\sigma^{j+}\gamma_5$.
 In the case of spin-0 target,  only the contribution of the Boer-Mulders
 functions can exist.
 The full list of leading-twist TMDs is shown in a schematic way in
 Tab.~\ref{t:TMDandIPD}. 
 
The IPDs are distribution functions  in the mixed momentum and coordinate space $(x,\vet{b}_\perp)$, with $\vet{b}_\perp$ being the transverse distance of the quark from the transverse center of momentum of the target~\cite{Burkardt:2000za}.
They
are defined from the following quark-quark correlator
\begin{equation}
\mathcal{F}^{[\Gamma]}(x,\vet{b}_\perp,S)=\frac{1}{2}\int \frac{dz^-}{2\pi}e^{ixp^+z^-}
\braket{p^+,\vet{R}_\perp=\vet{0}_\perp,S|\overline{\psi}(z_1)\Gamma
\mathcal{W}\ta z_1,z_2\tc \psi\ta z_2\tc|p^+,\vet{R}_\perp=\vet{0}_\perp,S},\label{IPDCORR}
\end{equation}
   where the quark fields are evaluated at  $z_{1,2}=(0^+,\mp\frac{z^-}{2},\vet{b}_\perp)$ and the hadron is in a state with longitudinal momentum $p^+$ at a transverse position
   $\vet{R}_\perp=\vet{0}_\perp$~\cite{Burkardt:2000za,Burkardt:2002hr,Soper:1976jc}.

The GPDs in the momentum space are defined through the following light-cone
correlation function
\begin{equation}
   F^{[\Gamma]}(x,\xi,t,S)= \frac{1}{2}\int \frac{dz^-}{2\pi} e^{ik\cdot
     z}\braket{p',S|\overline\psi\ta -\frac{z}{2}\tc \Gamma
     \mathcal{W}\ta-\frac{z}{2},\frac{z}{2}\tc \psi\ta
     \frac{z}{2}\tc|p,S}\Big|_{z^+=0, \ 
\vet{z}_\perp=\vet{0}_\perp}\label{GPDCORR}
\end{equation}
and  depend, beside $x$, on the following variables
\begin{align} 
\xi&=-\frac{\Delta^+}{2P^+},
&
t&=\Delta^2,
\end{align} 
where $P=(p+p')/2$ and $\Delta=p'-p$.
For $\xi=0$, the GPD correlator~\eqref{GPDCORR}  is related to the IPD correlator~\eqref{IPDCORR} by a Fourier transform from the coordinates $\vet{\Delta}_\perp$ to $\vet{b}_\perp$, and, accordingly, we can obtain the IPD ${\cal X}$ from the following Fourier transform of the GPD
\begin{equation} 
   {\cal X}(x,\vet{b}^2_\perp)=\int\frac{d\vetp{\Delta}}{(2\pi)^2}e^{-i\vet{\Delta}_\perp\cdot\vet{b}_\perp} X(x,\xi=0,-\vet{\Delta}^2_\perp).\label{FT-GPD}
\end{equation}     
 The full list of leading-twist IPDs is shown in a schematic way in
 Tab.~\ref{t:TMDandIPD}. 
At leading-twist and  for spin $1/2$ targets, the  correlator~\eqref{IPDCORR}  
with $\Gamma=\gamma^+$ and  transversely polarized targets  can be parametrized in terms of the
 derivative of the  IPD $\mathcal{E}$, while
with $\Gamma=i\sigma^{j+}\gamma_5$ and unpolarized target we access the derivative of the combination $\mathcal{E}_T +2\widetilde{\mathcal{H}}_T$ of  chiral-odd IPDs.
In the case of spin-zero targets, the contributions from the IPDs $\mathcal{E}$
and $\mathcal{E}_T $ are absent.

\renewcommand{\arraystretch}{1.5}
\begin{table}
\begin{center}
\begin{tabular}{c|c|c|c|c|}
\multicolumn{2}{c}{}&\multicolumn{3}{c}{quark pol.}\\
\cline{2-5}
& & U & L & T \\ \cline{2-5}
\multirow{3}{*}{\rotatebox{90}{nucleon pol.}} &{U} & ${\color{black} f_1}$   &
& $h_1^{\perp}$ \\
\cline{2-5}
& {L} & &${\color{black} g_1}$ &  $h_{1L}^{\perp}$ \\
\cline{2-5}
& {T} & $f_{1T}^{\perp}$  & $g_{1T}$ &  {\color{black} $h_1$}, $h_{1T}^{\perp}$ \\
\cline{2-5}
\multicolumn{1}{c}{}&\multicolumn{4}{c}{Twist-2 TMDs}
\end{tabular}
\qquad
\begin{tabular}{c|c|c|c|c|}
\multicolumn{2}{c}{}&\multicolumn{3}{c}{quark pol.}\\
\cline{2-5}
& & U & L & T \\ \cline{2-5}
\multirow{3}{*}{\rotatebox{90}{nucleon pol.}} &{U} & ${\color{black} \mathcal{H}}$
& & $\mathcal{E}_T+2 \widetilde{\mathcal{H}}_T$ \\
\cline{2-5}
& {L} & &${\color{black} \widetilde{\mathcal{H}} }$ & \\
\cline{2-5}
& {T} & $\mathcal{E}$  &   &
      {\color{black} $\mathcal{H}_T$},$\widetilde{\mathcal{H}}_T$ \\
\cline{2-5}
\multicolumn{1}{c}{}&\multicolumn{4}{c}{Twist-2 IPDs}
\end{tabular}
\end{center}
\caption{Tables of the leading-twist transverse momentum distributions (TMDs)
  and impact parameter distributions (IPDs) with their relation to nucleon and
  quark polarization states. For the complete definition, we refer to
  Refs.~\cite{Diehl:2005jf,Meissner:2007rx}.} 
\label{t:TMDandIPD}
\end{table}

The analogy  between the tensor structure of the parametrizations of the quark TMD and IPD correlators suggests the following correspondences for the distributions of spin $1/2$ targets~\cite{Diehl:2005jf,Meissner:2007rx}
\begin{align}
f_{1T}^\perp\ta x,\vet{k}^2_\perp \tc \leftrightarrow -\ta \mathcal{E}\ta x,\vet{b}^2_\perp\tc\tc',
\quad h_1^\perp\ta x,\vet{k}^2_\perp\tc \leftrightarrow -\ta \mathcal{E}_T\ta x,\vet{b}^2_\perp\tc+2\widetilde{\mathcal{H}}_T\ta x,\vet{b}^2_\perp\tc\tc',\label{eq:relations-nucleon}
\end{align}
where we used the following notation for the derivative of the IPDs
\begin{align}
\ta{\cal X}\ta x,\vet{b}^2_\perp\tc\tc'=\frac{\partial}{\partial \vet{b}^{2}_{\perp}}{\cal X}\ta x,\vet{b}^{2}_\perp\tc.
\end{align}
Similarly, the correspondence for spin-zero targets reads
\begin{align}
\quad h_1^\perp\ta x,\vet{k}^2_\perp\tc \leftrightarrow -\ta \widetilde{\mathcal{H}}_T\ta x,\vet{b}^2_\perp\tc\tc'.\label{eq:relations-pion}
\end{align}

In order to specify the precise form of the link
in Eqs.~\eqref{eq:relations-nucleon}, we
consider the average quark transverse momentum of an unpolarized quark in a
transversely polarized target given by
\begin{equation} 
 \braket{k_\perp^i(x)}_{UT}=\int d\vet{k}_\perp k^i_\perp
   \Phi^{[\gamma^+]}(x,\vet{k}_\perp, \vet{S}_\perp).
 \label{ktransv-sivers}
\end{equation} 

Following the derivation in Ref.~\cite{Meissner:2007rx},
Eq.~\eqref{ktransv-sivers} can be rewritten as
\begin{align} 
      \begin{split} 
 \braket{k_\perp^i(x)}_{UT}&=
 \frac{1}{2}\int \frac{dz^-}{2\pi} e^{ixp^+z^-} \braket{p,\vet{S}_\perp|\overline \psi\ta -\frac{z}{2}\tc 
\mathcal{W}\ta-\frac{z}{2},\frac{z}{2}\tc \mathcal{I}^{i}\ta\frac{z}{2}\tc  \gamma^+
 \psi\ta \frac{z}{2}\tc|p,\vet{S}_\perp}\big|_{z^+=\vetp{z}=0}
 \\
 &=
 \frac{1}{2}\int d\vet{b}_\perp 
 \int\frac{dz^-}{2\pi}
e^{ixp^+z^-}
\braket{p^+,\vet{R}_\perp=\vet{0}_\perp,\vet{S}_\perp| 
\bar{\psi}(z_1)\mathcal{W}(z_1;z_2)\mathcal{I}^{i}(z_2)\gamma^+\psi(z_2)|p^+,\vet{R}_\perp=\vet{0}_\perp,\vet{S}_\perp}
\label{ktransv-sivers-2}
\end{split} 
\end{align} 
The operator $\mathcal{I}^{i}(z)$ encodes the contribution of the FSIs, and is defined
as
\begin{equation}
  \begin{split} 
  \mathcal{I}^{i}\ta z\tc
  &= \frac{g_s}{2}\int dy^-\mathcal{W}\ta (z^-,z^+,\vetp{z}),(y^-,z^+,\vetp{z})\tc G^{+i}\ta y^-,z^+,\vetp{z}\tc\mathcal{W}\ta (y^-,z^+,\vetp{z}),(z^-,z^+,\vetp{z})\tc,
  \label{coordinateLensing}
  \end{split} 
\end{equation}
with $G^{+i}$ being the gluon-field strength tensor.

As observed for the first time in Ref.~\cite{Boer:2003cm}, there is a connection between
the transverse-momentum weighted quark correlator in Eq.~\eqref{ktransv-sivers-2} and the collinear
twist-3 quark-gluon-quark correlator defined as\footnote{In the literature there are various slightly different forms of the
parametrization of the quark-gluon-quark correlator~\cite{Boer:1997bw,Kanazawa:2000hz,
  Eguchi:2006mc,Boer:2003cm,Kanazawa:2014tda}. Here we use
the version of Ref.~\cite{Kanazawa:2015ajw}.}

\begin{align}
i \Phi_G^i(x,x) &=
 \int \frac{d z^- d y^-}{(2 \pi)^2}\,
e^{i xp^+ z^-}\,
\langle p,\vetp{S}| \overline{\psi}\ta-\frac{z}{2}\tc\, 
 \mathcal{W}\ta -\frac{z}{2},y \tc \, 
i g\, G^{+ i}(y) \, 
 \mathcal{W}\ta y,\frac{z}{2}\tc \,  
\psi\ta\frac{z}{2}\tc |p,\vetp{S} \rangle 
\Big|_{\substack{
z^+=\vetp{z}=0 \\
y^+=\vetp{y}=0}}
\\
&=  \frac{M}{2}\biggl[  -i \epsilon_\perp^{ij} S_\perp^j
  F_{FT}(x,x)
  \gamma^- 
 + H_{FU}(x,x)\, \gamma^i \gamma^- \biggr]\ .
\label{e:phiF} 
\end{align}
The decomposition contains the
so-called Qiu-Sterman matrix
element $F_{FT}$~\cite{Qiu:1991pp} and an analogous, chiral-odd term.\footnote{$F_{FT}$ introduced in
 \cite{Kanazawa:2015ajw} corresponds to $-T_F/(2M\pi)$ in \cite{Qiu:1991pp}.}

Eq.~\eqref{coordinateLensing} has a more intuitive interpretation
 in the
 light-cone gauge, $A^+ = 0$~\cite{Burkardt:2003yg}. In this case,
 the Wilson lines in the definition of $\mathcal{I}^{i}(z)$ run along the light-cone, and
 reduce to unity. As a result, one has
\[
\mathcal{I}^{i}\ta z\tc = \frac{g_s}{2}\Bigl( A_\perp^i\ta\infty^-,z^+,\vetp{z}\tc - A_\perp^i\ta-\infty^-,z^+,\vetp{z}\tc\Bigr).
\]
The gauge, however, is not completely fixed by the condition $A^+=0$. The fixing of residual gauge degrees of freedom can be obtained using additional boundary conditions on the gauge potential.
There are three common choices:
\begin{align*}   
  \vetp{A}(\infty^-) &= 0,
  &
  \vetp{A}(-\infty^-) &= 0,
  &
  \vetp{A}(\infty^-) +\vetp{A}(-\infty^-) &= 0,
\end{align*}     
known, respectively, as retarded, advanced and principal value prescription. 
We work with the advanced boundary condition  $\vetp{A}(-\infty^-) = 0$, but analogous results hold for the other two prescriptions (as it should be, since all the results must be gauge invariant). Our choice leads to the following results
\begin{align} 
 \braket{k_\perp^i(x)}_{UT}&=\frac{g_s}{2}\int \frac{dz^-}{2\pi} e^{ixp^+z^-} \braket{p,\vet{S}_\perp|\overline \psi\ta -\frac{z}{2}\tc A_\perp^i\ta \infty^-\tc\gamma^+
   \psi\ta \frac{z}{2}\tc|p,\vet{S}_\perp}\big|_{z^+=\vetp{z}=0},
 \\
\braket{k_\perp^i(x)}_{TU}^j& = \frac{g_s}{2}\int \frac{dz^-}{2\pi} e^{ixp^+z^-} \braket{p|\overline \psi\ta -\frac{z}{2}\tc A_\perp^i\ta \infty^-\tc i\sigma^{j+}\gamma_5\psi\ta \frac{z}{2}\tc|p}\big|_{z^+=\vetp{z}=0}.
\label{averageTransverseMomSingleGluon}
\end{align} 
One notices from Eq.~\eqref{averageTransverseMomSingleGluon} that  the FSIs in
the light-cone gauge with advanced boundary conditions (and, similarly, with
the retarded or principal value prescriptions) reduce to the exchange of a
transverse gluon at light-cone infinity between the active quark and the
spectator partons.

Up to this point, the analysis is still general.
In order to obtain an
expression containing an IPD, some very specific
conditions have to be imposed on the operator $\mathcal{I}^{i}(z)$,
Using  the completeness relation, we can rewrite  the first line of Eq.~\eqref{ktransv-sivers-2} as
\begin{align}
  \braket{k_\perp^i(x)}_{UT}  &= \frac{1}{2}\int d\vet{b}_\perp 
 \int\frac{dz^-}{2\pi}
 e^{ixp^+z^-}
 \SumInt_{X,X'}  \braket{X| \mathcal{I}^{i}(z_2)|X'}\nonumber
 \\
 &\quad \times
\braket{p^+,\vet{R}_\perp=\vet{0}_\perp,\vet{S}_\perp| 
  \bar{\psi}(z_1)\mathcal{W}(z_1,z_2) |X}
\gamma^+
\braket{X'|\psi(z_2)|p^+,\vet{R}_\perp=\vet{0}_\perp,\vet{S}_\perp}.\label{intermediate-states}
\end{align}
If we introduce the Fourier transform of the quark fields $\psi(z/2)$ and $
\phi\ta\frac{z}{2}\tc = \bar{\psi}\ta -\frac{z}{2}\tc\mathcal{W}\ta -\frac{z}{2};\frac{z}{2} \tc$, 
and use the light-front Fock expansion for the intermediate states, Eq.~\eqref{intermediate-states} reads as
\begin{align}
  \braket{k_\perp^i(x)}_{UT}   &= \frac{1}{2}   \int \{dk_1\} \{dk_2\} \{dl\}\int\frac{dz^-}{2\pi} e^{ixp^+z^-} e^{-i\frac{z^-}{2}(k^+_1+k^+_2+l^+)}\sum_{n,m}\sum_{\beta,\beta'} \int \prod_{i=1}^n \frac{dq^+_id\bm{q}_{\perp,i}}{(2\pi)^3 2q_i^+}  \prod_{i=1}^m \frac{dw^+_i d\bm{w}_{\perp,i}}{(2\pi)^3 2w_i^+}\nonumber\\
 &\times \braket{p^+,\vet{p}_\perp=\vetp{0},\vet{S}_\perp| \phi(k_1)\gamma^+ |\{q^+_i,\bm{q}_{\perp,i}\}_n}  \braket{\{q^+_i,\bm{q}_{\perp,i}\}_n,\beta'| I^{i}(l)|\{w^+_i,\bm{w}_{\perp,i}\}_m} \nonumber\\
 &\times\braket{\{w^+_i,\bm{w}_{\perp,i}\}_m,\beta'| \psi(k_2)|p^+,\vet{p}_\perp=\vetp{0},\vet{S}_\perp},\label{general1}
\end{align}
where $\{dk\} $ is the Lorentz invariant integration measure.
In Eq.~\eqref{general1}, the index $\beta$ and $\beta'$ label the parton,  color  and the helicity content of  the intermediate states.
As derived explicitly in the Appendix, one can obtain the factorization of the lensing function and the IPD in Eq.~\eqref{general1} by requiring that  the matrix elements of the operator $I^i(l)$ satisfy the following relation
\begin{equation}
 \braket{\{q^+_i,\bm{q}_{\perp,i}\}_n|
   I^{i}(l)|\{w^+_i,\bm{w}_{\perp,i}\}_m}  = 2\pi
 L^i\biggl(\frac{\vetp{l}}{1-x} \biggr) \delta_{n,m}\delta_{\beta\beta'}\delta(l^+)\prod_{i=1}^n (2\pi)^32q_i^+\delta(q^+_i-w^+_i)\delta\ta \bm{q}_{\perp,i} - \bm{w}_{\perp,i} -x_i\frac{\vetp{l}}{1-x}\tc,
 \label{IMatEl}
\end{equation}
where $x_i$ is the light-cone momentum fraction of each constituent w.r.t. the hadron target light-cone momentum, i.e. $x_i=w^+_i/p^+$, and should satisfy the relation $\sum_i x_i=1-x$.
The matrix elements of the operator $I^i(l)$ in Eq.~\eqref{IMatEl} represent the  interaction between the active parton and the spectator system mediated by the Wilson gluons and 
correspond to the FSIs that occur in a SIDIS process.
The relation~\eqref{IMatEl} imposes strict conditions that
are equivalent to requiring that:
\begin{itemize}
\item[1)]  the FSIs should connect Fock states with the same number of constituents and the same parton, helicity and color content; 
\item[2)] the FSIs should transfer the total transverse momentum  $\vetp{l}/(1-x)$ to the whole spectator system;
\item[3)] the FSIs can not transfer momentum in the light-cone direction to the spectator system;
\item[4)] the FSIs should transfer a fraction $x_i=w^+_i/p^+$ of the total transverse momentum  to each constituent of the spectator system.
\end{itemize}

The last condition is the most stringent. It is crucial  to obtain the correct transverse light-front boost that gives the nondiagonal matrix element defining the GPD and then  the transverse distortion in impact parameter space described by the IPD.  
For convenience, we can discuss the implications of the condition $4)$  
in the light-cone gauge with advanced boundary conditions, where the FSI reduces to the exchange of a transverse gluon at light-cone infinity between the active parton and the spectator system.
In this case, one can easily deduce that the condition $4)$ can  be realized with a perturbative coupling between the gauge boson and the active parton only if the spectator system 
  is composed by a single constituent, i.e. the hadron target is a two-body bound system.  Then, the light-cone momentum fraction of the spectator is equal to $1-x$ and 
  the constraint on the transverse momentum transferred by the Wilson gluon to the spectator system follows trivially from the conservation of the total momentum of the hadron target. 
  Otherwise, the condition $4)$ imposes to share the transverse momentum
  carried by the Wilson gluon with each spectator parton in a proportion equal
  to the longitudinal momentum fraction $x_i$. This can not be realized in systems composed by more than two constituents by assuming an interaction vertex between the gauge boson and a single constituent.
 
We conclude that
if and only if the above conditions are fulfilled we can write
\begin{equation} 
\begin{split}
  \braket{k_\perp^i(x)}_{UT}
  &=
  -\int d\vet{k}_\perp k^i_\perp\frac{\epsilon_\perp^{jk}k^j_\perp S^k_\perp}{M}
  f^\perp_{1T}(x,\vet{k}^2_\perp)
  \\
  & \approx\int d\vet{b}_\perp
         \mathcal{L}^{i} \bigl(\vet{b}_\perp/(1-x)\bigr)
         \mathcal{F}^{[\gamma^+]}(x,\vet{b}_\perp,\vet{S}_\perp)
  \\ &
         = \int d\vet{b}_\perp\mathcal{L}^{i}\bigl(\vet{b}_\perp/(1-x)\bigr)\frac{\epsilon^{jk}_\perp b^j_\perp S^k_\perp}{M}
   \left( \mathcal{E}(x,\vet{b}^2_\perp)\right)'.
         \label{relation-sivers3}
  \end{split}
\end{equation} 
 In the next sections, we will consider explicitly a few model calculations and we will discuss to which extent the conditions 1) -- 4) can be satisfied.
 
In an analogous way, we can analyze the
    average quark transverse momentum of a transversely polarized quark in an unpolarized target given by
\begin{equation} 
 \braket{k_\perp^i(x)}^j_{TU}=\int d\vet{k}_\perp k^i_\perp
 \Phi^{[i\sigma^{j+}\gamma_5]}(x,\vet{k}_\perp, S)
 .
 \label{ktransv-bm}
\end{equation}
With similar steps as before, under the conditions of applicability of the
lensing hypothesis, we obtain
\begin{equation}
  \begin{split} 
 \braket{k_\perp^i(x)}^j_{TU}&=-\int d\vet{k}_\perp
 k^i_\perp\frac{\epsilon_\perp^{kj}k^k_\perp
 }{M}h^\perp_{1}(x,\vet{k}^2_\perp)
 \\
 &\approx
\int d\vet{b}_\perp \mathcal{L}^{i}\bigl(\vet{b}_\perp/(1-x)\bigr)\mathcal{F}^{[i\sigma^{j+}\gamma_5]}(x,\vet{b}_\perp)
\\
&= \int d\vet{b}_\perp\mathcal{L}^{i}\bigl(\vet{b}_\perp/(1-x)\bigr)\frac{\epsilon^{kj}_\perp b^k_\perp}{M} \Bigl(\mathcal{E}_T(x,\vet{b}^2_\perp)
 +2\widetilde{\mathcal{H}}_T(x,\vet{b}^2_\perp)\Bigr)'.\label{relation-bm3}
  \end{split}
\end{equation}

Alternatively, by contracting Eqs.~\eqref{relation-sivers3} and
\eqref{relation-bm3}  with
$-\epsilon_\perp^{il} S^l_\perp/(2M)$ and $-\epsilon_\perp^{ij}/(2M)$, respectively, we
can write
\begin{align} 
  f_{1T}^{\perp(1)}(x)& = \pi F_{FT}(x,x)
  \approx
  \frac{1}{4}\int d\vet{b}_\perp b_{\perp}^i\mathcal{L}^{i}\bigl(\vet{b}_\perp/(1-x)\bigr)
  \mathcal{E}^{(1)}(x,\vet{b}^2_\perp),\label{relation-sivers4}
 \\
 h_{1}^{\perp(1)}(x)& = \pi H_{FU} (x,x)
 \approx\frac{1}{4}\int d\vet{b}_\perp
 b_{\perp}^i\mathcal{L}^{i}\bigl(\vet{b}_\perp/(1-x)\bigr)
      \left(\mathcal{E}_T^{(1)}(x,\vet{b}^2_\perp)
      +2\widetilde{\mathcal{H}}_T^{(1)}(x,\vet{b}^2_\perp)\right),
\label{relation-bm4}
\end{align}
where we used the following notations
\begin{align}
  f^{(1)}(x,\vet{k}_\perp^2)
    &= \frac{\vet{k}^2_\perp}{2M^2}
  f(x,\vet{k}_\perp^2)
  ,  
  \\
{\cal X}^{(1)}(x, \vet{b}^2_\perp) &= -\frac{2}{M^2}
\frac{\partial}{\partial \vet{b}^{2}_{\perp}} {\cal X} (x, \vet{b}^2_\perp)  =
   \int\frac{d\vetp{\Delta}}{(2\pi)^2}e^{-i\vet{\Delta}_\perp\cdot\vet{b}_\perp}
   \frac{\vet{\Delta}^2_\perp}{2M^2} X(x,\xi=0,-\vet{\Delta}^2_\perp).
\label{eq:f-der1}
\end{align} 
For spin-zero targets, only Eqs.~\eqref{relation-bm3} and \eqref{relation-bm4} with  $\mathcal{E}_T(x,\vet{b}^2_\perp)=0$ and $2\widetilde{\mathcal{H}}_T\rightarrow \widetilde{\mathcal{H}}_T.$
survives.

\section{Lensing relation for the pion}
\label{PionLensingFunction} 
In this section, we show how the relation between the Boer-Mulders function and the the chiral-odd GPD $\widetilde{\mathcal{H}}_T$ is realized for a spin-zero target like the pion, described as a   quark-antiquark ($q\bar q$) bound state.
In the framework of light-front quantization and working in the gauge $A^+ = 0$, the leading-order contribution of the Fock-state decomposition of a pion state with momentum $p$  is given by
\begin{equation}
\ket{\pi(p)}  = \sum_{\{\lambda_i\}} \sum_{\{q_i\}} \int \qa Dx\qc_2 \Psi_{q\bar q}(\beta,r)\delta_{c_1c_2}\ket{\lambda_1,q_1,c_1,p_1}\ket{\lambda_2,q_2,c_2,p_2}.
\label{LFWFpion}
\end{equation}
In Eq.~\eqref{LFWFpion}, $\lambda_i$ are the quarks light-front helicities, $q_i = q, \bar q$ denotes the quark and anti-quark flavor, $c_i$ is a color index, and $p_i$ is the parton momentum. The function $\Psi_{q\bar q}$ is the light-front wave function (LFWF) of the $q\bar q$ state and its arguments are 
indicated with the collective notation  $\beta = (\{\lambda_i\},\{q_i\})$ and $r = \{x_i,\vet{k}_{\perp,i}\}$.  The momentum coordinates $k_i$ of the partons  are in the 
so-called ``hadron" frame~\cite{Diehl:2000xz}, corresponding to the reference frame where the pion has zero transverse momentum, i.e.
\begin{align}
\bm{k}_{\perp,i} &= \bm{p}_{\perp,i} - x_i\vetp{p}, \quad  x_i= \frac{p_i^+}{p^+} = \frac{k_i^+}{p^+}.\label{intrinsic-coordinates}
\end{align}
We will  refer to transverse parton momenta in the hadron frame as intrinsic transverse momenta.
The integration measure in Eq.~\eqref{LFWFpion} is defined as
\begin{align}
\qa Dx\qc_N &= \frac{[dx]_N[d\vetp{k}]_N}{\sqrt{\prod_{i=1}^N x_i}}, \label{tot-measure-int} \\
[dx]_N &= \delta \ta 1-\sum_{i=1}^Nx_i\tc \prod_{i=1}^N dx_i,\label{long-measure-int} \\
[d\vetp{k}]_N &= 2(2\pi)^3\delta \ta \sum_{i=1}^N\bm{k}_{\perp,i}\tc \prod_{i=1}^N \frac{d\bm{k}_{\perp,i}}{2(2\pi)^3}. \label{transv-measure-int}
\end{align}
The flavor and helicity structure of the parton composition in Eq.~\eqref{LFWFpion} can be made explicit as~\cite{Ji:2003yj}
\begin{equation}
  \begin{split} 
&\ket{\pi(p)} = T_\pi \int \frac{dx_1dx_2}{\sqrt{x_1x_2}} \frac{d\bm{k}_{\perp,1}d\bm{k}_{\perp,2}}{2(2\pi)^3} \delta\ta 1-x_1-x_2\tc \delta\ta \bm{k}_{\perp,2}+\bm{k}_{\perp,1}\tc \frac{\delta_{c_1c_2}}{\sqrt{3}} \\
& \quad \times \Bigl\{ \psi^{(1)}(1,2)\qa q^{c_1\dagger}_{\uparrow}(1)\bar q^{c_2\dagger}_{\downarrow}(2) - q^{c_1\dagger}_{\downarrow}(1)\bar q^{c_2\dagger}_{\uparrow}(2)\ket{0}\qc 
+ \psi^{(2)}(1,2)\qa \bm{k}_{L,1} q^{c_1\dagger}_{\uparrow}(1)\bar q^{c_2\dagger}_{\uparrow}(2) +  \bm{k}_{R,1} q^{c_1\dagger}_{\downarrow}(1)\bar q^{c_2\dagger}_{\downarrow}(2)\ket{0}\qc\Bigr\},
\label{pionState}
  \end{split}
\end{equation}
where $\vet{k}_{R(L),i}=k_{x,i}\pm ik_{y,i}$ and $q^{c_i\dagger}_{\lambda}$ and $\bar{q}^{c_i\dagger}_\lambda$ are the creation operators of quark and antiquark with helicity $\lambda$ and color $c_i$, respectively.
In Eq.~\eqref{pionState}, $T_\pi$ is the isospin factor which projects on the different members of the isotriplet of the pion, and is defined as $T_\pi = \sum_{\tau_q,\tau_{\bar q}} \braket{1/2\tau_q 1/2\tau_{\bar q} |1\tau_\pi}$ with $\tau_{q,\bar q, \pi}$ the isospin of the quark, anti-quark and pion state, respectively.
Furthermore, the  the light-front wave amplitudes (LFWAs) $\psi^{(1)}$ and $\psi^{(2)}$  are functions of quark momenta with arguments $1,2$ representing $x_1$, 
$\vet{k}_{\perp, 1}$ and $x_2$, $\vet{k}_{\perp, 2}$, respectively, and correspond  to quark states with orbital angular momentum (OAM) $l_z=0$ and $|l_z|=1$, respectively. They are scalar functions, and depend on the parton momenta only through scalar products $\vet{k}_{\perp,i}\cdot\vet{k}_{\perp,j}$.
In the light-cone gauge with  advanced boundary conditions for the transverse components of the gauge field, the LFWAs  are complex functions~\cite{Ji:2002aa,Belitsky:2002sm,Brodsky:2010vs}.
Using the pion state~\eqref{pionState}, we can represent the pion GPDs and TMDs in terms of  overlap of LFWAs in a model-independent way.

The pion chiral-odd GPD  is defined as
\begin{align}
F^{[i\sigma^{j+}\gamma_5]}_\pi(x,\Delta^+,\vet{\Delta}_\perp)&=-\frac{i\epsilon_\perp^{kj}\Delta^k_\perp}{M_\pi} \widetilde{H}_{T,\pi}(x,\xi,-\vet{\Delta}_\perp^2),\label{eq:gpd-correlator-pion}
\end{align}
where $\epsilon_\perp^{ij}=\epsilon^{+-ij}$.
Introducing   the following overlap of LFWAs for the $q\bar q$ component of the pion 
\begin{align}
G^k\ta 1,1'\tc &= F^k\bigl( x_1,\bm{k}_{\perp,1};1-x_1,-\bm{k}_{\perp,1}\bigr| \bigl| x'_1,\bm{k}'_{\perp,1}; 1-x'_1,-\bm{k}'_{\perp,1}\bigr),\\
F^k\ta(1,2 \db 1',2'\tc) & = k_{\perp,1}^{k}\psi^{(2)}(1,2)\psi^{(1)*}(1',2') - k'^k_{\perp,1}\psi^{(1)}(1,2)\psi^{(2)*}(1',2'), 
\label{GpionDef}
\end{align}
  one finds at $\xi=0$
\begin{align}
\frac{\Delta_{\perp}^k}{2M_\pi} \widetilde{H}_{T,\pi}(x,0,-\vet{\Delta}_\perp^2)
= \frac{T_\pi^2}{2(2\pi)^3} \int d\vetp{k} G^k\ta x, \vetp{k}  \db x, \vetp{k} + (1-x)\vetp{\Delta}\tc.
\label{H1piDefXi}
\end{align}
Note that the Wilson line in the light-cone correlator defining the GPDs formally reduce to  the identity in the light-cone gauge $A^+=0$.  
We can Fourier transform the integral in Eq.~\eqref{H1piDefXi}, with the result
\begin{equation}
  \begin{split} 
\int d\vetp{k} G^k\ta x, \vetp{k}  \db x, \vetp{k} + (1-x)\vetp{\Delta}\tc & = \int d\vetp{k}  \int d\vetp{A}d\vetp{B} e^{-i\vetp{A}\cdot\vetp{k} + i\vetp{B}\cdot(\vetp{k} + (1-x)\vetp{\Delta})}\mathcal{G}^k\ta x , \vetp{A}  \db x, \vetp{B}\tc \\
& = \int d\vetp{B} e^{i(1-x)\vetp{B}\cdot\vetp{\Delta}}\mathcal{G}^k\ta x , \vetp{B}  \db x, \vetp{B}\tc.
\end{split}
\end{equation} 
Using this expression, Eq.~\eqref{H1piDefXi} can easily be  transformed into the impact parameter $\vetp{b}$ space to obtain the pion chiral-odd IPD 
\begin{equation}
  \begin{split} 
\frac{ib_{\perp}^k}{M_\pi}\ta \widetilde{\mathcal{H}}_{T,\pi}(x,\vet{b}_\perp^2) \tc'
&= \int \frac{d\vetp{\Delta}}{(2\pi)^2} e^{-i\vetp{b}\cdot\vetp{\Delta}}\left( \frac{\Delta_{\perp}^k}{2M_\pi}H_{T,\pi}(x,0,-\vet{\Delta}_\perp^2)  \right) \\
& = \frac{T_\pi^2}{2(2\pi)^3}\int \frac{d\vetp{\Delta}}{(2\pi)^2} e^{-i\vetp{b}\cdot\vetp{\Delta}}\int d\vetp{B} e^{i(1-x)\vetp{B}\cdot\vetp{\Delta}}\mathcal{G}^k\ta x, \vetp{B}  \db x, \vetp{B}\tc  \\
& = \frac{T_\pi^2}{2(2\pi)^5(1-x)^2}\mathcal{G}^k\ta x , \frac{\vetp{b}}{1-x}  \db x, \frac{\vetp{b}}{1-x}\tc .
  \end{split}
\end{equation}

\begin{figure}
\centering
\includegraphics[width = 0.5\textwidth]{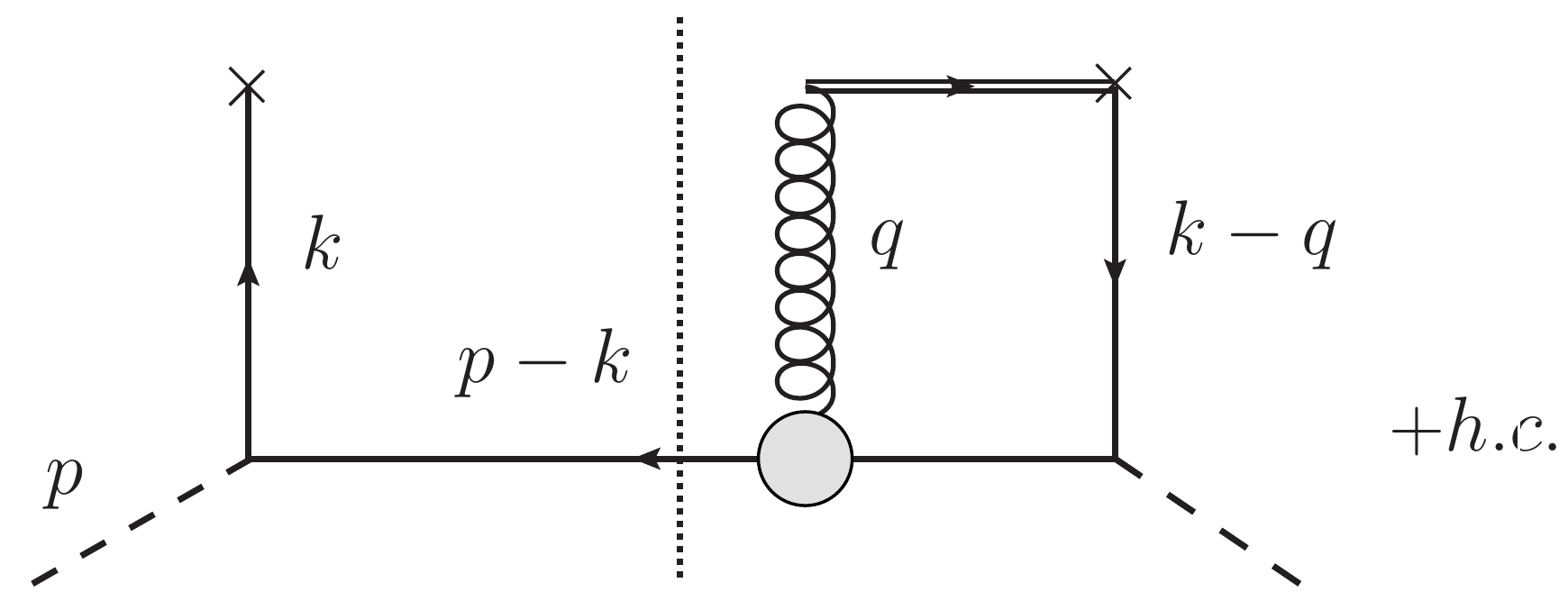}
\caption{Cut diagram contributing to the  average transverse momentum 
of T-odd effects in single-spin asymmetries of a SIDIS process, in the $A^+=0$
gauge and for a pion target in the  $q\bar q$ configuration.
}
\label{effCouplingDiagram}
\end{figure}

The same Dirac structure  giving  the GPD $\widetilde{H}_{T,\pi}$ in Eq.~\eqref{H1piDefXi}  enters the  correlator 
that defines the  Boer-Mulders TMD, i.e.
\begin{equation}
\Phi^{[i\sigma^{j+}\gamma_5]}_\pi = -\frac{ \epsilon_\perp^{kj}k^k_{\perp}}{M_\pi} h_{1,\pi }^{\perp}(x,\vet{k}_\perp^2).
\label{piTMDdef}
\end{equation}
The tensor structures in Eqs.~\eqref{eq:gpd-correlator-pion} and~\eqref{piTMDdef} 
have opposite behavior under time reversal, which reveals the 
T-even and T-odd nature of $\widetilde{H}_{T,\pi}$ and $h_{1,\pi}^\perp$, respectively.
This crucial difference reflects on the fact that the Boer-Mulders function would be zero without the contribution of the Wilson line.
In the light-cone gauge, the contribution along the light-cone direction vanishes, and there remains a residual 
contribution
 from the transverse link at $\xi^- = \infty^-$.
As outlined in Sec. \ref{TheoreticalRelations}, in the calculation of the average transverse momentum of the Boer-Mulders effect the Wilson line reduces to the exchange of one Wilson gluon  between the active quark and the spectator system (see Eq.~\eqref{averageTransverseMomSingleGluon}).
If we describe the pion as a bound $q\bar q$ system, the corresponding cut diagram 
 can be represented as in
Fig. \ref{effCouplingDiagram}, where the blob indicates an effective coupling
between the antiquark and the Wilson gluon 
 We will assume that the coupling is perturbative and consider only the tree level graph.
 In this framework, 
the LFWA overlap representation of the Boer-Mulders  function has been derived   in Ref.  \cite{Pasquini:2014ppa} and reads 
 \begin{equation}
k_{\perp}^kh_{1,\pi}^\perp\ta x,\vetp{k}^2\tc = \frac{2\alpha_s}{(2\pi)^4} \frac{4}{3}T_\pi^2 M_\pi \int \frac{d\vetp{q}}{\vetp{q}^2} G^k\ta x, \vetp{k}\db x,\vetp{k}-\vetp{q}\tc ,
\label{BMexpr}
\end{equation}
where $\vetp{q}$ is the transverse momentum of the Wilson gluon (we recall that $q^+ = 0$ in the eikonal approximation in the light cone gauge).
We note that  Eq.~\eqref{BMexpr} involves the same overlap of LFWAs as in Eq.~\eqref{H1piDefXi} for the case of the GPD $\widetilde{H}_{T,\pi}$.
With the formal identification of 
\begin{equation}
-\vetp{q} = (1-x)\vetp{\Delta},
\label{momRelPi}
\end{equation}
the $G^k$ functions in Eqs.~\eqref{H1piDefXi} and~\eqref{BMexpr} have the same momentum dependence. This is crucial to recover the lensing function relation. Using Eq.~\eqref{BMexpr}, we can now calculate the average transverse momentum of the Boer-Mulders effect 
\begin{equation}
  \begin{split} 
\braket{k_{\perp}^i}^j_{TU} & = -\int d\vetp{k} k_{\perp}^i \frac{\epsilon_\perp^{kj}k^k_\perp}{M_\pi}
h_{1,\pi}^\perp = -\frac{2\alpha_s}{(2\pi)^4} \frac{4}{3}T_\pi^2
 \int \frac{d\vetp{q}}{\vetp{q}^2} \int d\vetp{k}  k_{\perp}^i \epsilon_\perp^{kj}G^k\ta x, \vetp{k}\db x,\vetp{k}-\vetp{q}\tc  \\
& =-\frac{2\alpha_s}{(2\pi)^4} \frac{4}{3}T_\pi^2 
\int\frac{d\vetp{q}}{\vetp{q}^2} \int d\vetp{k}k_{\perp}^i \epsilon_\perp^{kj}  \int d\vetp{A}d\vetp{B} e^{-i\vetp{A}\cdot\vetp{k}}e^{i\vetp{B}\cdot(\vetp{k} - \vetp{q})}
\mathcal{G}^k\ta x, \vetp{A}\db x,\vetp{B}\tc  \\
& =  -i\frac{4}{3}4\pi\alpha_s (1-x)^3\int d\vetp{B} \int\frac{d\vetp{q}}{\vetp{q}^2} \frac{\epsilon_\perp^{kj}B_{\perp}^k}{M_\pi} q_{\perp}^i e^{-i\vetp{B}\cdot\vetp{q}}  \ta \widetilde{\mathcal{H}}_{T,\pi}(x,\vet{B}_\perp^2(1-x)^2)\tc',
\label{LensingDerivation}
  \end{split}
\end{equation}
where we used the following relation
\begin{equation}
\begin{split}
\int d\vetp{k}k_{\perp}^i  & \int d\vetp{A}d\vetp{B} e^{-i\vetp{A}\cdot\vetp{k}}e^{i\vetp{B}\cdot(\vetp{k} - \vetp{q})}\mathcal{G}^k\ta x, \vetp{A}\db x,\vetp{B}\tc \notag 
=\frac{q^i_\perp}{2}\int d\vetp{B} e^{-i\vetp{B}\cdot\vetp{q}}{\cal G}^k(x,\vetp{B}||x,\vetp{B})
\nonumber\\
&=iq^i_\perp(1-x)^3\frac{(2\pi)^5}{M_\pi T^2_\pi}\int d\vetp{B} B^k_\perp\left(\widetilde{\mathcal{H}}_{T,\pi}(x,\vetp{B}^2(1-x)^2))\right)'e^{-i\vetp{B}\cdot\vetp{q}}.
\end{split}
\end{equation}
With the variable change $\vet B_\perp\rightarrow \vet{b}_\perp/(1-x)$ in Eq.~\eqref{LensingDerivation}, we then find 
\begin{equation}
\braket{k_{\perp}^i}^j_{TU} =\int d\vetp{b} \frac{\epsilon_\perp^{kj}b^k_\perp}{M_\pi} \mathcal{L}^{i}\ta  \vetp{b}/(1-x)\tc \ta \widetilde{\mathcal{H}}_{T,\pi}(x,\vet{b}_\perp^2) \tc',
\label{lensing-pion-final}
\end{equation}
where we introduced the lensing function~\cite{Burkardt:2003je}
\begin{equation} 
\mathcal{L}^{i}\ta \vetp{b}/(1-x)\tc =- i\frac{4}{3}\alpha_s 4\pi  \int \frac{d\bm{q}_{\perp}}{\vetp{q}^2} q_{\perp}^ie^{-i\frac{\vetp{b}\cdot \vet{\bm{q}}_{\perp}}{(1-x)}}=
-\frac{8}{3}\alpha_s 4\pi^2 \frac{b^{i}_\perp }{\vet{b}_\perp^2}(1-x).
\label{LensingFunctionTemp}
\end{equation}

A few comments are in order. This result relies on the assumption that  the
coupling between the Wilson gluon and the spectator parton is perturbative,
i.e. it is described by the tree-level QCD vertex. The coupling, at leading
power of $1/Q\simeq 1/p^+$, conserves the helicity of the spectator
parton. Therefore, the helicity flip of the active quark must be compensated
by a change of the OAM carried by the partons  in
the initial and final states. This situation is equivalent to the GPD case,
where the active and spectator quarks do not change the helicity and the
helicity flip of the target must be compensated by a transfer of OAM between
the partons in the initial and final states. 

Due to the two-body nature of the problem ($q \bar q$ system) the role of the
transverse momentum of the gluon $\vetp{q}$ is the same as the external
transverse momentum $\vetp{\Delta}$ in the GPD case. This can be traced back
to the fact that the parton distributions should be invariant by light-front
transverse boosts 
and 
depend on the intrinsic transverse-momentum coordinates of the partons.
In the case of the average transverse-momentum of the Boer-Mulders effect,
there is no change of the transverse momentum of the pion between the initial
and final state. However,  the quark and anti-quark have a different intrinsic
transverse momentum in the initial and final states due to the gluon
exchange. 
 In the GPD case, the momentum transferred to the pion is absorbed by the
 active quark, while the transverse-momentum of the spectator quark does not
 change in the initial and final states.  In terms of the intrinsic transverse
 momentum coordinates~\eqref{intrinsic-coordinates} in the hadron-in and
 hadron-out frame of the initial and final hadrons, respectively, 
both the active and spectator quarks experience a transfer of transverse
momentum. 
Therefore, one can make the formal identification of Eq.~\eqref{momRelPi} in the momentum dependence of the LFWAs describing the contribution of 
the internal parton dynamics and  the effect of the FSIs can be factorized in
the lensing function. 
In the next section, we will see that in the case of a three body system this
correspondence can not be established, and, as a consequence, the
lensing-function relation breaks down. 

If we do not assume a perturbative coupling between the Wilson gluon and the
anti-quark spectator, we may have an effective vertex that depends on the
momenta of the gluon and of the anti-quark. 
The coupling with the Wilson gluon may occur with or without flip of the
helicity of the anti-quark. 
In the first case, the lensing relation can not hold, as it will be discussed
in Sec. \ref{ProtonModels}.  
 If the helicity flip is not allowed, the lensing relation can still  be
 spoiled by the dependence of the vertex on the momentum of the anti-quark. 
By introducing an effective vertex with a general parametrization
$\Lambda\ta \vetp{k}^2, \vetp{q}^2, \vetp{k}\cdot\vetp{q}\tc \gamma^+$, the average transverse momentum of the Boer-Mulders effect in
Eq.~\eqref{LensingDerivation} becomes
\begin{align}
\braket{k_{\perp}^i}^j_{TU}& =-\frac{2\alpha_s}{(2\pi)^4} \frac{4}{3}T_\pi^2 
\int d\vetp{k} k_{\perp}^i\epsilon_\perp^{kj}\int \frac{d\vetp{q}}{\vetp{q}^2}
G^k\ta x, \vetp{k}\db x,\vetp{k}-\vetp{q}\tc \Lambda\ta \vetp{k}^2, \vetp{q}^2, \vetp{k}\cdot\vetp{q}\tc
\notag\\
& = -\frac{2\alpha_s}{(2\pi)^4} \frac{4}{3}T_\pi^2 
\Bigg\{
\int d\vetp{B}d\vetp{A}\int \frac{d\vetp{q}}{\vetp{q}^2}  e^{-i\vetp{B}\cdot\vetp{q}}\frac{q_{\perp}^i}{2} \epsilon_\perp^{kj}\mathcal{G}^k\ta x, \vetp{B}-\vetp{A}\db x, \vetp{B}\tc \widetilde\Lambda\ta \vetp{A}^2, \vetp{q}^2, \vetp{A}\cdot\vetp{q}\tc  \notag \\
&\quad   + \frac{i}{2}\int d\vetp{B}d\vetp{A}\ \int \frac{d\vetp{q}}{\vetp{q}^2}e^{-i\ta \vetp{B} + \vetp{A}\tc \cdot \frac{\vetp{q}}{2}}\epsilon_\perp^{kj}  \notag \\
&\quad   \times \partial_{A_{\perp,i}} \qa \mathcal{G}^k\Big( x, \vetp{A}\Big|\Big| x,\frac{\vetp{B}+\vetp{A}}{2}\Big)  \widetilde \Lambda\ta \ta\frac{\vetp{B}-\vetp{A}}{2}\tc^2,\vetp{q}^2, \frac{\vetp{B}\cdot\vetp{q} - \vetp{A}\cdot\vetp{q}}{2}\tc\qc \Bigg\}.
\label{eq:vertex-eff}
\end{align}

In Eq.~\eqref{eq:vertex-eff}, it is impossible to factorise a term like $\mathcal{G}\ta x, \vetp{B}\db x,\vetp{B}\tc$ and recognize the definition of the IPD, as done in Eq.~\eqref{LensingDerivation}. 
However, if  the effective vertex depends only on $\vetp{q}$, then the lensing relation~\eqref{lensing-pion-final} still holds, with the following  new definition of the lensing function
\begin{equation} 
\mathcal{L}^{i}\ta\vetp{b}/(1-x)\tc =- i\frac{4}{3}\alpha_s 4\pi  \int \frac{d\bm{q}_{\perp}}{\vetp{q}^2} q_{\perp}^i\Lambda(\vetp{q}^2)
e^{-i\frac{\vetp{b}\cdot \vet{\bm{q}}_{\perp}}{(1-x)}}.
\label{LensingFunction-effective}
\end{equation}

\section{Lensing relation for the proton}
\label{ProtonLensing}
In this section we discuss  the validity of the lensing relations in Eq.~\eqref{eq:relations-nucleon} for the proton system.
For illustration,
we will consider in detail the relation between the Sivers TMD and the IPD $\mathcal{E}$. However,
 the same arguments can be applied for the relation involving the Boer-Mulders TMD and the combination $\mathcal{E}_T+2\widetilde{\mathcal{H}}_T$ of chiral-odd IPDs

We  limit ourselves to analyzing the general structure of the LFWA overlap representation of the GPDs and the Sivers function, since the explicit dependence on the LFWAs is not relevant for our purposes. We refer to~\cite{Boffi:2002yy,Pasquini:2010af} for the full calculation.
We introduce the  LFWAs overlap 
$ F_T\bigl( x_1,\bm{k}_{\perp,1}; x_2,\bm{k}_{\perp,2}; x_3,\bm{k}_{\perp,3} \bigr|\bigl| x'_1,\bm{k}'_{\perp,1}; x'_2,\bm{k}'_{\perp,2}; x'_3,\bm{k}'_{\perp,3} \big) 
$
and define the function $G_T$ as
\begin{align*}
& G_T\ta x_1,\bm{k}_{\perp,1}; x_2,\bm{k}_{\perp,2} \big|\big| x'_1,\bm{k}'_{\perp,1}; x'_2,\bm{k}'_{\perp,2}\tc \\
& \ = F_T\ta x_1,\bm{k}_{\perp,1}; x_2,\bm{k}_{\perp,2}; 1-x_1-x_2,-\bm{k}_{\perp,1}-\bm{k}_{\perp,2} \big|\big| x'_1,\bm{k}'_{\perp,1}; x'_2,\bm{k}'_{\perp,2}; 1-x'_1-x'_2,-\bm{k}'_{\perp,1}-\bm{k}'_{\perp,2} \tc,
\end{align*}
where the arguments on the right-hand side of $||$ refer to the momentum dependence  of the complex conjugate LFWF of the proton in the final state, and the arguments on the left-hand side give the momentum dependence of the LFWF of the proton in the initial state.

The GPD $E$ in the limit of $\xi = 0$ 
is obtained from the quark-quark correlator~\eqref{GPDCORR} with $\Gamma=\gamma^+$ and transversely polarized proton.
The incoming and outgoing 
quark momenta are related by $p'_i= p_i$  $(i\ne j)$ for the spectator quarks and $p'_j= p_j + \Delta$ for the active quark that takes the momentum transfer to the proton.
The intrinsic momenta are then obtained via the transverse-boost in Eq.~\eqref{intrinsic-coordinates} and are related as
\begin{align}
  x'_i&=x_i,
  &
  \vet{k}'_{\perp,i} &= \vet{k}_{\perp,i} - x_i\vetp{\Delta},
  & \text{for } i\ne j,\\
  x'_j&=x_j,
  &
  \vet{k}'_{\perp,j} &= \vet{k}_{\perp,j} +(1-x_j)\vetp{\Delta}.
\end{align}
Using momentum conservation for the intrinsic variables, i.e., $\sum_i x_i=1$
and $\sum_i \vet{k}_{\perp,i}=\vetp{0}=\sum_i \vet{k}'_{\perp,i}$, one finds
the following LFWF overlap representation~\cite{Boffi:2002yy}
\begin{equation} 
\frac{i\epsilon_\perp^{ij}\Delta^j_\perp S_T^i}{M}E(x,\xi=0,-\vetp{\Delta}^2) = \frac{1}{4(2\pi)^6}\int d\vetp{k} \int_{0}^x dy\int d\vetp{t} G_T\ta  x,\vetp{k}; y,\vetp{t}\db x, \vetp{k}+ (1-x)\vetp{\Delta};y,\vetp{t}-y\vetp{\Delta}\tc.
\label{GPDEdef}
\end{equation} 
The results for the IPD distribution are then obtained taking the Fourier transform  of Eq.~\eqref{GPDEdef}  w.r.t.~$\vetp{b}$ and expressing $G_T$ in terms of  its Fourier integral. One finds

\begin{equation}
-\frac{\epsilon_\perp^{ij}b_{\perp}^jS_T^i}{M}\ta \mathcal{E}(x,\xi=0,\vetp{b}^2)\tc'= \frac{1}{4(2\pi)^8}\frac{1}{1-x} \int_{0}^x dy \int d\vetp{B} \mathcal{G}_T\ta x, \frac{y\vetp{B}-\vetp{b}}{1-x}; y, \vetp{B} \db x, \frac{y\vetp{B}-\vetp{b}}{1-x}; y, \vetp{B}\tc.
\label{IPDE}
\end{equation}

The LFWF overlap representation of the Sivers function has been derived in  Ref. \cite{Pasquini:2010af},  using the $3q$ component of the nucleon state and the one-gluon exchange approximation, with a perturbative quark-gluon coupling. 
It is given by the same function  $G_T$  as for the GPD $E$, but with
different arguments, i.e.,
\begin{equation} 
\frac{\epsilon_\perp^{ij}k^j_\perp S_T^i}{M} f_{1T}^\perp \ta x, \vetp{k}^2\tc=
-\frac{ \alpha_s}{3(2\pi)^7}\int \frac{d\vetp{q}}{\vetp{q}^2} \int_0^x dy
\int d\vetp{t} G_T\ta  x,\bm{k}_{\perp}; y,\bm{t}_{\perp}\db x,
\vetp{k}-\vetp{q};y,\vetp{t}+\vetp{q}\tc. 
\label{siversOGE}
\end{equation}
From this expression, one clearly sees that the formal identification in Eq.~\eqref{momRelPi} does not apply in the case of Eqs.~\eqref{GPDEdef}
 and~\eqref{siversOGE}, since $(1-x)$ and $y$ are independent variables.
As we will see, this is sufficient to break the lensing-function relation in the case of the proton.

From Eq.~\eqref{siversOGE}, one can calculate the  average transverse momentum
of the Sivers effect as 
\begin{equation}
  \begin{split} 
\braket{k_{\perp}^i}_{UT} & = -\int d\vetp{k} k_{\perp}^i\frac{\epsilon_\perp^{ij}k^j_\perp S_T^i}{M} f_{1T}^\perp  \\
& =  \alpha_s\frac{M}{3(2\pi)^7}\int d\vetp{k} k_{\perp}^i\int \frac{d\vetp{q}}{\vetp{q}^2} \int_0^x dy\int d\vetp{t} G_T\ta  x,\bm{k}_{\perp}; y,\bm{t}_{\perp}\db x, \vetp{k}-\vetp{q};y,\vetp{t}+\vetp{q}\tc  \\
& =\alpha_s\frac{M}{3(2\pi)^7} \int_0^x dy\int d\vetp{A}d\vetp{B} \int \frac{d\vetp{q}}{\vetp{q}^2} \frac{q_{\perp}^i}{2}e^{i\vetp{q}\cdot\vetp{A}} \mathcal{G}_T\ta x, \vetp{B}-\vetp{A}; y, \vetp{B} \db x, \vetp{B}-\vetp{A}; y, \vetp{B}\tc  \\
& = 
-i \alpha_s\frac{M}{6(1-x)(2\pi)^6}\
\int_0^x dy\int d\vetp{b}d\vetp{B} \frac{b_{\perp}^i}{\vetp{b}^2}
\mathcal{G}_T\ta x, \frac{\vetp{B}-\vetp{b}}{1-x}; y, \vetp{B} \db x,\frac{
  \vetp{B}-\vetp{b}}{1-x}; y, \vetp{B}\tc .
  \end{split}
\end{equation}

Comparing this equation with Eq.~\eqref{IPDE}, we immediately notice that the different dependence of the function $G_T$ on $\vetp{B}$ does not allow us to
 factorize the contribution of the IPD from a lensing function.
This can be traced back to the fact that in the LFWF overlap representation of the GPD the transverse momentum $\vetp{\Delta}$ appears multiplied by both $(1-x)$ and $y$, 
since both the two spectator quarks have different intrinsic transverse momentum in the initial and final states. In other words,
the  transverse boost~\eqref{intrinsic-coordinates}  from a given frame to the hadron frames transforms the transverse-momentum coordinates of the two spectator quarks in a different way, depending on their fraction  $x_i$ of longitudinal momentum.
Vice versa, in the TMD case the hadron does not change the transverse momentum in the initial and final states, 
and the gluon interaction 
occurs between the active quark and a single spectator quark, leaving unchanged the intrinsic momentum of the other spectator quark.

The failure of the lensing relation is ultimately due to the three-body structure of the nucleon LFWF, and
persists more in general for any hadron system described by more than two constituent partons.
Therefore the lensing relation is spoiled also for the pion when considering Fock-state components beyond the leading-order $q\bar{q}$ state.
For the nucleon, one may recover the lensing relation  in models where the nucleon is described as a two-body system, such as the  models with a quark and a diquark spectator.
However,
one has to distinguish between different variants of diquark-spectator models, depending on the spin structure of the diquark and its coupling with the Wilson gluon, as we will discuss in the following section.

\section{Diquark spectator models for the nucleon}
\label{ProtonModels}
The basic idea of spectator models is to evaluate the quark-quark correlators
entering the definition of the TMDs and of the GPDs by inserting a complete
set of intermediate states and then truncating this set at tree level to a
single on-shell spectator diquark state, i.e., a state with the quantum numbers
of two quarks. The diquark can be either an isospin singlet with spin 0
(scalar diquark) or an isospin triplet with spin 1 (axial-vector diquark). The
target is then seen as made of an off-shell quark and an on-shell
diquark. Spectator models differ by their specific choice of the
target-quark-diquark vertex, of the polarization four-vectors associated with
the axial-vector diquark, and of the vertex form factor that takes into
account the composite nature of the target   in an effective way. The
approximation of a diquark spectator spoils part of the richness of the
nonperturbative structure of the proton, that cannot be captured by the
vertex form factor alone. Moreover, some models may introduce relations (like
the lensing relation, but not only) that do not hold in general and are due to
the simplifications introduced by the models themselves. A review of
model-induced relations between different TMDs and between TMDs and GPDs can
be found in Refs.~\cite{Meissner:2008ay,Avakian:2010br,Lorce:2011zta}.  

The lensing relations hold in the scalar spectator models, as it was first
shown in Ref.~\cite{Burkardt:2003uw}. The arguments which lead to the lensing
relations are essentially the same as discussed in
Sec.~\ref{PionLensingFunction} for the pion, i.e., the hadron 
described as 
a two-body system and the assumption of a perturbative helicity-conserving
coupling between the gauge boson and the spectator system. 
For the axial-vector diquark model (AVDQ), the validity of the lensing-function relations  depends on the helicity-structure of the diquark.
One way to classify AVDQ models is to distinguish between models that allow
for the presence of a longitudinal polarization and models that admit only
transverse polarizations for the axial-vector diquark. 
The first class corresponds to the AVDQ models that can not satisfy the
lensing relations,  in contrast to the second  group.

\begin{figure}[t!]
\begin{center}
\includegraphics[width = 0.5\textwidth]{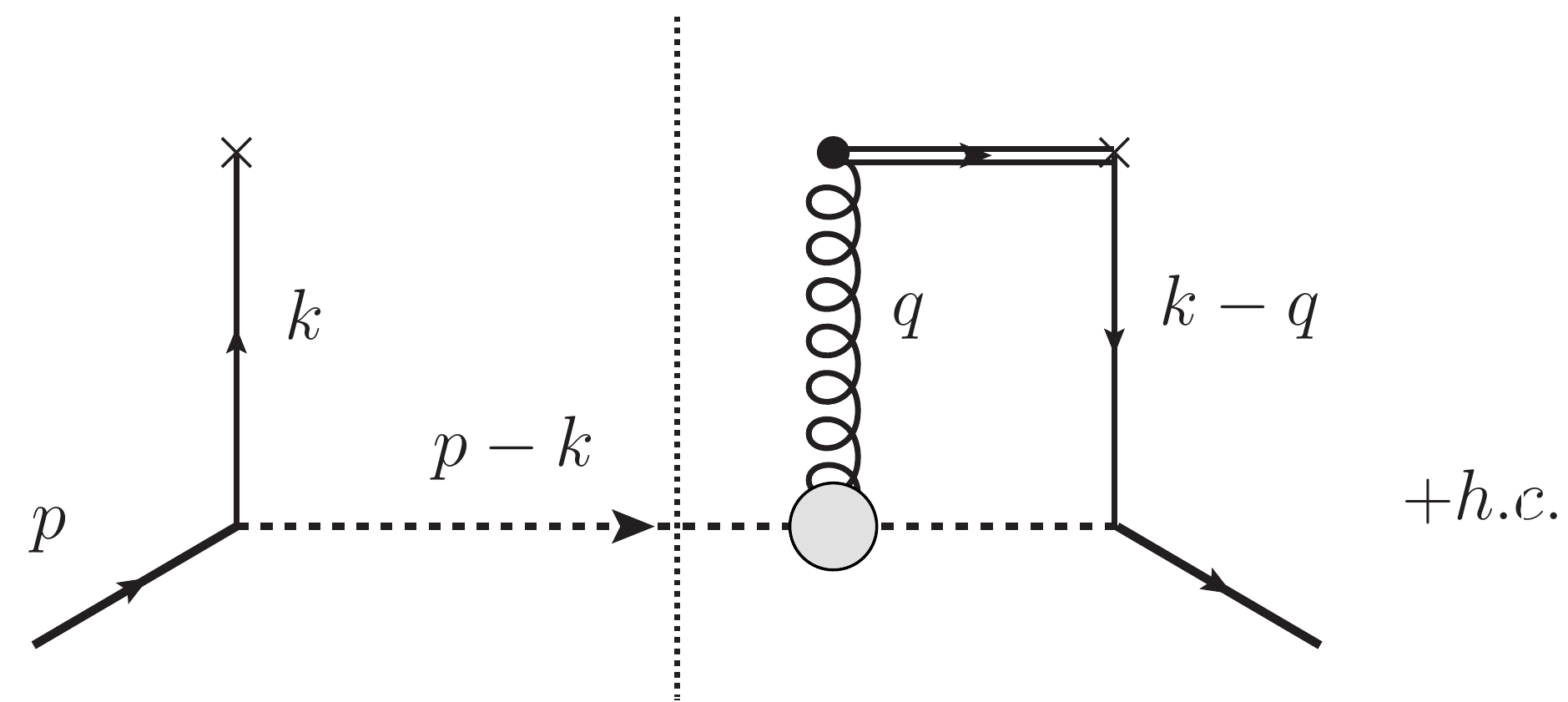}
\end{center}
\caption{Cut diagram contributing to the average transverse-momentum of T-odd effects in single spin asymmetries of a SIDIS process, in the $A^+=0$ gauge and for a proton target 
described in a quark-diquark model.}
\label{axialDiquark}
\end{figure}

To illustrate this, we introduce the  polarization vectors of the AVDQ (see,
e.g., Ref.~\cite{Bacchetta:2008af}): 
\begin{align}
\varepsilon_{+1}(l) &= \ta 0, \frac{-l_R}{\sqrt{2}l^+}, {\bm \varepsilon}_{+1,\perp}\tc, \label{AVDQ-1} \\
\varepsilon_{-1}(l) &= \ta 0, \frac{-l_L}{\sqrt{2}l^+}, \bm{\varepsilon}_{-1,\perp}\tc ,\label{AVQD-2}\\
\varepsilon_{0}(l) &= \frac{1}{M_a}\ta l^+, \frac{l_\perp^2 - M_a^2}{2l^+}, \vetp{l}\tc,\label{AVDQ-3}
\end{align}
where $M_a$ is the AVDQ mass, $l \equiv p-k$, and:
\[
\bm{\varepsilon}_{+1,\perp} = -\ta \bm{\varepsilon}_{-1,\perp}\tc^* = -\frac{1}{\sqrt{2}}\ta 1,i \tc .
\]
Here we do not consider the (unphysical) time-like polarization that is discussed in Ref. \cite{Bacchetta:2008af}.
The polarization vectors in Eqs.~\eqref{AVDQ-1}-\eqref{AVDQ-3} satisfy the following relations:  $\varepsilon_{+1}(l)\cdot\varepsilon^*_{-1}(l') = 0$ for any value of $l,l'$, whereas 
$\varepsilon_{\pm 1}(l)\cdot\varepsilon^*_{0}(l) = 0$ 
and $\varepsilon_{\pm 1}(l)\cdot\varepsilon^*_{0}(l') \ne 0$ for $l\ne l'$.
The interaction between the diquark and the gluon is given by the following   coupling tensor
\begin{equation}
\frac{i}{e_c}\Gamma_{\nu\sigma}^{\rho} = (2l+q)^\rho g_{\nu\sigma} - (l+(1+\kappa_a)q)_\sigma\delta^\rho_\nu - (l-\kappa_aq)_\nu\delta^\rho_\sigma,\label{gluon-vertex}
\end{equation}
where  $e_c$ and $\kappa_a$ are, respectively,   the diquark  color charge  and  the diquark anomalous chromomagnetic moment, which takes into account that 
 the diquark is not a point-like massive axial particle, but is an effective constituent degree of freedom. 
In the calculation of the  T-odd TMDs, the indices of the coupling tensor~\eqref{gluon-vertex} are saturated with  the gluon propagator and the AVDQ polarization vector (see Fig.~\ref{axialDiquark}). The contraction of the  coupling tensor  with the polarization vectors of the axial-vector diquark gives the following interaction vertex
\begin{equation}
\mathcal{R}^{\rho}= 
\sum_{\lambda_1,\lambda_2 = \pm 1,0}\varepsilon^{\nu*}_{\lambda_1}(l) \varepsilon^{\sigma}_{\lambda_2}(l+q) \Gamma_{\nu\sigma}^{\rho}.\label{vertex-avdq}
\end{equation}
This expression can be compared with the corresponding $\bar q g \bar q$ interaction vertex, which enters the calculation of the Boer-Mulders function of the pion, i.e.
\begin{equation}
\mathcal{R}^{\rho} =
\sum_{\lambda_1,\lambda_2 = \pm 1/2}\bar v_{\lambda_1}(l+q)\gamma^\rho v_{\lambda_2}(l).\label{vertex-pion}
\end{equation}
In both cases, the vertex function has the following scaling behavior
\begin{align}
  \mathcal{R}^{+} &\simeq \mathcal{O}(p^+),
  &
  \mathcal{R}^{i}_\perp&\simeq \mathcal{O}(1),
  &
\mathcal{R}^{-} &\simeq \mathcal{O}(1/p^+).
\end{align}
However, in the case of the pion, the leading-order term $\mathcal{R}^{+}$ in
Eq.~\eqref{vertex-pion} is helicity conserving, whereas the leading
contribution $\mathcal{R}^{+}$ in Eq.~\eqref{vertex-avdq} for the axial-vector
diquark  contains terms that flip the helicity of the diquark.  
As a result, the following transitions are allowed for the AVDQ  interacting with the Wilson gluon
\[
\lambda_a = \pm 1 \leftrightarrow \lambda_a = 0, \pm 1.
\]
In the calculations of the GPDs, the Wilson line in the light-cone gauge reduces to unity, and 
the spectator can not flip the helicity between the initial and final states. 
We conclude that the LFWA overlaps must be different for the average  transverse momentum of the T-odd TMD functions and for the GPDs, 
hence the lensing relation cannot hold. 
Only  if one assumes that the longitudinal polarization for the AVDQ is absent in the proton, 
the lensing relation can be restored.
This situation occurs within the  quark target model, where the  nonAbelian  three gluon vertex enters the computation of the T-odd TMDs and allows only for helicity-conserving transitions.

\section{Conclusions}
\label{sect:conclusions}

In this work, we have investigated the origin of nontrivial relations between
transverse distortions in the distribution of quarks in impact-parameter space
and analogous distortions in transverse-momentum space. The former are encoded
in impact-parameter distributions (IPDs) and contribute to observable
asymmetries in exclusive processes involving hadrons. The latter are
encoded in the T-odd Sivers and 
Boer-Mulders transverse-momentum distributions (TMDs),
and give rise to observable
asymmetries in semi-inclusive processes involving hadrons.

We have identified the conditions under which
it is possible to express the Sivers and
Boer-Mulders functions as convolutions of an IPD and a lensing function,
incorporating the effects of the FSIs between the active parton and the rest
of the hadron.
These conditions, listed in Sec.~\ref{TheoreticalRelations},
appear to be very specific and
hold in a restricted class of models.

To better illustrate the nature of these conditions,
we checked the validity or failure of the lensing relation in three models:
(1) a model of the pion described as a
quark-antiquark bound state, (2) a model of the proton as a bound state of three
quarks, and (3) a model of the proton as a bound state of a quark and a
spectator.
The conditions of validity of the lensing relation are more easily fulfilled in
models where the hadron is described as a two-body bound system, as in models
(1) and (3). However, they can be violated even in these simple models, as
happens in certain versions of (3), e.g.,
with axial-vector spectators that admit longitudinal polarization.
Finally,
the conditions are violated in model (2) and in general can not be 
obtained if the hadron is described by more than two constituents and the interaction vertex of the gauge boson occurs with a single  constituent.

In conclusion, it seems that the lensing relation is
unlikely to survive in the full complexity of nonperturbative QCD, even
approximately. Phenomenological studies of the Sivers and Boer-Mulders
function, as well as possible lattice QCD studies, should be able to confirm
the violation of the lensing hypothesis.

\section*{Acknowledgments}
This work is supported by the the European Union's Horizon 2020 programme
under grant agreement No. 824093 (STRONG2020) and under the European
Research Council (ERC) grant agreement No. 647981 (3DSPIN).
\appendix
\section{}

In this appendix we explicitly  derive the conditions 1) -- 4) on the matrix
elements of the FSI operator  discussed in
Sec.~\eqref{TheoreticalRelations}.   
Condition 1) follows from the requirement that the IPD  we want to factorize  in Eq.~\eqref{general1} is diagonal in the parton Fock space.
Analogously, condition 2) is necessary to recover the correct Fourier transform of the quark fields that enters the definition of the IPD correlator.
Conditions 3) and 4) are a consequence of momentum conservation.
  The matrix element of the function $I^i(l)$ in Eq.~\eqref{IMatEl} connects states with total momenta given by
\begin{align}
\vetp{W}=\sum_{i=1}^n \vet{w}_{\perp, i}, &\quad W^+=\sum_{i=1}^n w^+_i,\nonumber\\
\vetp{Q}=\sum_{i=1}^n \vet{q}_{\perp, i}, &\quad Q^+=\sum_{i=1}^n q^+_i.\label{total-momenta}
\end{align}
By imposing total momentum conservation in each matrix elements of
Eq.~\eqref{general1},  we have
\begin{equation} 
\vetp{Q} = \vetp{W} + \vetp{l},\quad Q^+=W^+=(1-x)p^+.\label{momentum-cons}
\end{equation} 
Eqs.~\eqref{total-momenta} and \eqref{momentum-cons} are equivalent to
\begin{equation} 
\sum_{i=1}^n\vet{q}_{\perp,i} = \sum_{i=1}^n \vet{w}_{\perp,i} + \vetp{l}, \quad \sum_{i=1}^n \frac{q^+_i}{p^+}=\sum_{i=1}^n \frac{w^+_i}{p^+}=\sum_{i=1}^n x_i = 1-x.\label{total-relations}
\end{equation} 
Combining the two relations in Eq.~\eqref{total-relations}, we find
\begin{equation} 
\vet{q}_{\perp,i} =\vet{w}_{\perp,i} +\frac{x_i}{1-x} \vetp{l}.
\end{equation} 
As  final  result,  the conditions 1) -- 4) can be recast in the expression of
Eq.~\eqref{IMatEl} for  the matrix element of the lensing function. By
inserting Eq.~\eqref{IMatEl} in Eq.~\eqref{general1}, we have
\begin{equation} 
  \begin{split} 
 \braket{k_\perp^i(x)}_{UT} &= \frac{1}{2}   \int \{dk_1\} \{dk_2\}
 \frac{d\vetp{l}}{(2\pi)^2}\int\frac{dz^-}{2\pi} e^{ixp^+z^-}
 e^{-i\frac{z^-}{2}(k^+_1+k^+_2)}\sum_{n}\sum_\beta\int  \prod_{i=1}^m
 \frac{dw^+_i d\bm{w}_{\perp,i}}{(2\pi)^32w_i^+}
 \\
 &\quad \times L^i\ta \frac{\vetp{l}}{1-x}
 \tc\braket{p^+,\vet{0}_\perp,\vet{S}_\perp| \phi(k_1)\gamma^+
   |\left\{w^+_i,\bm{w}_{\perp,i}+x_i\frac{\vetp{l}}{1-x}\right\}_m}
 \braket{\{w^+_i,\bm{w}_{\perp,i}\}_m|
   \psi(k_2)|p^+,\vet{0}_\perp,\vet{S}_\perp}.
 \label{general2}
  \end{split}
\end{equation}
We now use the invariance of the matrix elements in Eq.~\eqref{general2} under transverse light-front boosts to obtain
\begin{align}
\braket{k_\perp^i(x)}_{UT} &= \frac{1}{2}   \int \{dk_1\} \{dk_2\} \frac{d\vetp{l}}{(2\pi)^2}\int\frac{dz^-}{2\pi} e^{ixp^+z^-} e^{-i\frac{z^-}{2}(k^+_1+k^+_2)}\sum_{n}\sum_\beta
 \int \prod_{i=1}^m \frac{dw^+_i d\bm{w}_{\perp,i}}{(2\pi)^32w_i^+} \nonumber\\
 &\quad \times L^i\ta \frac{\vetp{l}}{1-x} \tc\braket{p^+, -\vetp{l},\vet{S}_\perp| \phi(z_1)\gamma^+ |\{w^+_i,\bm{w}_{\perp,i}\}_m}  \braket{\{w^+_i,\bm{w}_{\perp,i}\}_m| \psi(z_2)|p^+,\vet{0}_\perp,\vet{S}_\perp} \nonumber\\
 &= \frac{1}{2}   \int\frac{d\vetp{l}}{(2\pi)^2}\int\frac{dz^-}{2\pi} e^{ixp^+z^-} L^i\ta\frac{\vetp{l}}{1-x}\tc\braket{p^+, -\vetp{l},\label{general3}
 \vet{S}_\perp| \phi(k_1)\gamma^+  \psi(k_2)|p^+,\vet{0}_\perp,\vet{S}_\perp}.
 \end{align}
 Eq.~\eqref{general3} can be finally Fourier transformed in the impact parameter space, with the result
 \begin{align}
\braket{k_\perp^i(x)}_{UT} &= \frac{1}{2}  \int \frac{d\vetp{l}}{(2\pi)^2}\int\frac{dz^-}{2\pi} e^{ixp^+z^-}  \int d\vetp{b} e^{-i\vetp{b}\vetp{l}} L^i\ta \frac{\vetp{l}}{1-x} \tc\braket{p^+,\vet{R}_\perp =\vetp{0} ,\vet{S}_\perp| \phi(z_1)\gamma^+  \psi(z_2)|p^+,\vet{R}_\perp =\vetp{0},\vet{S}_\perp}\nonumber\\
  &= \frac{1}{2} \int\frac{dz^-}{2\pi} e^{ixp^+z^-}  \int d\vetp{b}  \mathcal{L}^i\ta \frac{\vetp{b}}{1-x} \tc\braket{P^+,\vet{R}_\perp =\vetp{0} ,\vet{S}_\perp| \phi(z_1)\gamma^+  \psi(z_2)|P^+,\vet{R}_\perp =\vetp{0},\vet{S}_\perp},
\end{align}
where we recognize the convolution of the lensing function $\mathcal{L}(\vetp{b}/(1-x))$ and the correlator for unpolarized quark in a transversely  polarized target that is related  the IPD 
$\left(\mathcal{E}(x,\vetp{b}^2)\right)'$, i.e.
\begin{align}
\braket{k_\perp^i(x)}_{UT} &=  \int d\vetp{b}  \mathcal{L}^i\ta \frac{\vetp{b}}{1-x} \tc\frac{\epsilon_\perp^{jk}b^j_\perp S^k_\perp}{M}\left(\mathcal{E}(x,\vetp{b}^2)\right)'.
\end{align}

\section*{References}
%

\end{document}